\documentclass[aps,prl,reprint,10pt,longbibliography,superscriptaddress,nofootinbib]{revtex4-2}

\usepackage{amsmath, amssymb, amsfonts, amsthm, bm, mathrsfs, bbm, enumerate,mathtools}
\usepackage{graphicx}
\usepackage{epstopdf}
\usepackage{color}
\usepackage[usenames,dvipsnames]{xcolor}
\usepackage[citecolor=blue, colorlinks=true, urlcolor=blue, linkcolor=blue]{hyperref}
\usepackage{bbold}
\usepackage{feynmp-auto} 

\usepackage{soul}
\usepackage{float}

\usepackage{sansmath}
\DeclareFontEncoding{LGR}{}{}
\DeclareSymbolFont{sfgreek}{LGR}{cmss}{m}{n}
\SetSymbolFont{sfgreek}{bold}{LGR}{cmss}{bx}{n}
\DeclareMathSymbol{\sxi}{\mathord}{sfgreek}{`x}
\DeclareMathSymbol{\stheta}{\mathord}{sfgreek}{`j}
\DeclareMathSymbol{\sepsilon}{\mathord}{sfgreek}{`e}
\DeclareMathSymbol{\sOmega}{\mathalpha}{sfgreek}{`W}
\DeclareMathSymbol{\sdelta}{\mathalpha}{sfgreek}{`d}

\usepackage{color}
\usepackage{nicefrac}
\usepackage{dsfont} 
\usepackage{wasysym} 
\usepackage{stmaryrd} 
\usepackage{hyperref} 
\usepackage{xcolor}
\usepackage{empheq} 

\usepackage{verbatim}

\usepackage{cancel}
\usepackage{url}

\newcommand{\bs}{\boldsymbol}

\newcommand\varpm{\mathbin{\vcenter{\hbox{
  \oalign{\hfil$\scriptstyle+$\hfil\cr
          \noalign{\kern-.3ex}
          $\scriptscriptstyle({-})$\cr}
}}}}

\begin{document}

\newcommand{\ourtitle}{Detecting exciton condensation via charge transport in solid state Bose-Fermi mixtures }
\renewcommand{\ourtitle}{Detecting Exciton Condensation through Charge Transport in Semiconductor Heterostructures }

\title{\ourtitle}

\newcommand{\TUM}{\affiliation{Technical University of Munich, TUM School of Natural Sciences, Physics Department, 85748 Garching, Germany}}
\newcommand{\MCQST}{\affiliation{Munich Center for Quantum Science and Technology (MCQST), Schellingstr. 4, 80799 M{\"u}nchen, Germany}}

\author{Caterina Zerba} 
\TUM 
\MCQST

\author{L\'eo Mangeolle} 
\TUM 
\MCQST

\author{Michael Knap} 
\TUM 
\MCQST

\begin{abstract}

Direct evidence of exciton condensation in semiconductor heterostructures remains elusive.
Here we propose charge transport of doped carriers as a probe of exciton condensation in transition-metal dichalcogenide heterostructures and identify distinct experimental signatures.
First, condensation suppresses the phase space for carrier scattering, leading to a reduction in resistivity, that provides a general diagnostic of exciton condensation. 
Second, in heterostructures with a tunable solid-state Feshbach resonance, condensate-induced hybridization between doped carriers and trion bound states qualitatively modifies transport. 
In particular, near resonance, this hybridization yields a negative effective mass and a corresponding sign reversal of the Hall resistivity.
These results establish charge transport as a promising route for detecting and characterizing exciton condensation in semiconductor heterostructures.

\end{abstract}

\date{\today}

\maketitle

\textit{\textbf{Introduction --}}  
Excitons in semiconductor heterostructures have been theoretically proposed to offer an interesting platform for exploring Bose-Einstein condensation, due to favorable transition temperatures and rich multi-component order arising from the internal spin structure~\cite{keldysh1965, Zhu1995,Fogler2014, Wu2015, Fernandez-Rossier1997}. Transition metal dichalcogenide (TMD) heterostructures, in particular, combine strong Coulomb interactions with electrical tunability and support high density of tightly bound interlayer excitons, making them promising candidates for exciton condensation~\cite{Ma2021,Park2023,Xiong2023,Qi2023thermodynamic,Gao2024,Lian2024,Qi2025electrically,Nguyen2025,mhenni2024,Upadhyay2026,Qi2025drag,qi2026}.
Recent measurements of the spin susceptibility provide evidence consistent with condensation~\cite{qi2026}, although an experimentally accessible transport probe is still lacking. A hallmark would be counterflow measurements~\cite{Eisenstein2014}, in which oppositely directed currents flow in the two layers, but its observation is challenged by interlayer leakage and contacting. 

\begin{figure*}[!t]
\centering
\includegraphics[width=0.95\textwidth]{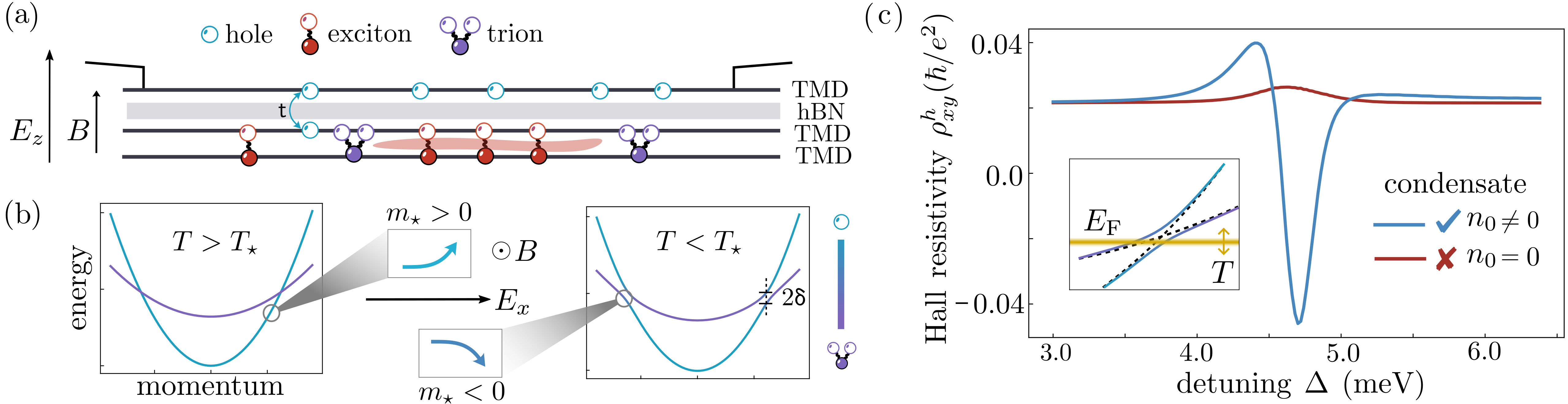}
\caption{\textbf{Setup and sharp transport signatures of exciton condensation.} (a) We consider a multilayer TMD structure; the upper layer is hole-doped, while interlayer excitons are electrically injected between the middle and lower layers. Holes tunnel from the upper to the middle layer and scatter with excitons, forming trions as a stable bound state. A transverse electric field $E_z$ tunes the scattering. Electric contacts with a reservoir induce a field $E_x$ in the upper layer which generates a hole current, and a transverse magnetic field $B$ realizes the Hall effect. (b) When excitons condense, $T<T^*$, the hole and trion bands hybridize. The local band curvature defines the local effective mass $m^\star$, which in turn determines the sign of the Hall effect (insets). (c) We consider fixed temperature $T$ and exciton density $n_x$: Without condensate ($n_0 \rightarrow 0$) the Hall resistivity $\rho^h_{xy}$ is mostly featureless, whereas with condensate ($n_0 \neq 0$) a two-peak structure  with a large \emph{negative} peak in $\rho^h_{xy}$ arises; a distinctive feature of exciton condensation. This is a direct consequence of hole-trion hybridization, which (see inset) is enhanced by a large hybridization gap, ${\sdelta}$, and low temperature, $T \ll {\sdelta}$.
}
\label{fig:1}
\end{figure*}

Here, we propose \emph{charge transport} of doped carriers as a probe of exciton condensation. TMD heterostructures naturally realize such Bose–Fermi mixtures~\cite{Qi2025electrically,Nguyen2025,mhenni2024}, where itinerant fermionic carriers coexist with bosonic excitons and their trion bound states. Interactions between these constituents are electric-field tunable by a solid-state Feshbach resonance~\cite{Kuhlenkamp21, schwartz21, Zerba2024, Zerba2025}. We identify the following transport signatures of condensation. First, in general condensation reduces the phase space available for carrier scattering, resulting in a change in resistivity upon cooling. Second, in heterostructures with a tunable solid-state Feshbach resonance, condensate-induced hybridization between carriers and trions  reconstructs the band structure; see Fig.~\ref{fig:1}. In the vicinity of the Feshbach resonance, this reconstruction generates a negative effective carrier mass and drives a sign reversal of the Hall resistivity, yielding a striking and tunable qualitative signature of the exciton condensate. Our theoretical analysis focuses on the more complex but also more tunable solid-state Feshbach setting. The resistivity change arising from the modified scattering phase space, however, provides a transport probe of exciton condensation beyond the specific system considered here.

\textit{\textbf{Setup --}} We consider a heterostructure composed of three TMD layers, as depicted in Fig.~\ref{fig:1} (a). The top layer is hole doped, and separated from the other two by an insulating layer of hexagonal boron nitride (hBN), that still enables hole tunneling to the middle layer with a tunneling rate $\sf t$~\cite{schwartz21}. In the middle and bottom layers, interlayer excitons are electrically injected. As these are tightly bound, they can be treated as independent well-defined bosonic excitations, that interact with holes tunneling from the top layer \cite{Kuhlenkamp21, schwartz21, Zerba2024, Zerba2025}. Because of this interaction, excitons and holes can form a fermionic ``trion" bound state, whose properties are tunable by a perpendicular electric field $E_z$. As in atomic physics~\cite{chin_rev_10}, tuning the bound state controls the effective scattering amplitudes of holes and excitons via the solid state-analogue of a Feshbach resonance \cite{Kuhlenkamp21}. 

This setting thereby realizes a strongly-interacting Bose-Fermi mixture, that is analytically tractable by treating the interactions to be mediated by the trion bound state \cite{schwartz21, Kuhlenkamp21, Zerba2024, Zerba2025}. This approach is exact at resonance, and allows us to define an effectively perturbative interaction vertex $\hat H_{\rm int}= g\int_{\bs p,\bs k}(\hat m^\dagger_{\bs{p+k}}\hat h_{\bs p}\hat x_{\bs k} + \rm h.c.)$ between holes ($\hat h^\dagger$), excitons ($\hat x^\dagger$) and trions ($\hat m^\dagger$). At relevant energy scales all three species have a quadratic dispersion, $\epsilon_h(\bs p)=p^2/2m_h-\mu_h$, $\epsilon_x(\bs k)=k^2/2m_x-\mu_x$ and $\epsilon_t(\bs p)=p^2/2m_t-\mu_t+ \Delta$, where $m_h,m_x,m_t$ are the hole, exciton and trion masses, and $\mu_h,\mu_x,\mu_t$ their chemical potentials, respectively. We also introduce the detuning $\Delta={\sf e}E_zd - E_t^0$, with ${\sf e}$ the hole charge, $d$ the hBN thickness, and $E_t^0$ the trion binding energy~\cite{Zerba2024,Zerba2025}. The different components of the Bose-Fermi mixture are in equilibrium, implying that the chemical potentials are related by $\mu_h+\mu_x=\mu_t$. We further assume the masses to satisfy $m_t/3=m_x/2=m_h$.

Our goal is to investigate the consequences of exciton condensation on the transport properties of itinerant holes of the Bose-Fermi mixture. We treat the exciton condensation at the mean-field level and recast the original exciton operators as 
$\hat x_{\bs k}^\dagger = \delta_{\bs k ,\bs 0} \sqrt{n_0}+  (1-\delta_{\bs k ,\bs 0})(\tilde u_{\bs k}\hat b^\dagger_{\bs k} - \tilde v_{\bs k}\hat b_{-\bs k})$,
with $\tilde u_{\bs k}^2 - \tilde v_{\bs k}^2 = 1$, where $n_0$ is the condensed fraction and the bosons $\hat b^\dagger_{\bs k}$ are now sound modes. 
The condensation of excitons thereby hybridizes trions and holes, $\hat H_{\rm int} \rightarrow \hat H_{\rm hyb} + \hat H_{\rm int}'$, where $\hat{H}_{\rm hyb} =  g\sqrt {n_0} \int_{\bs p} ( \hat{h}^\dagger_{\bs p} \hat{m}_{\bs p} + {\rm h.c } )$. The hybridization is diagonalized by introducing two new fermionic species $f_{\bs p}=u_{\bs p}m_{\bs p}-v_{\bs p}h_{\bs p}$ and $e_{\bs p}=v^*_{\bs p}m_{\bs p} + u^*_{\bs p}h_{\bs p}$ with a unitary transformation, $|u_{\bs p}|^2+|v_{\bs p}|^2=1$; the new bands exhibit a hybridization gap $2\sdelta = 2 g \sqrt{n_0}$ at the avoided crossing;  see Fig.~\ref{fig:1} (b). In this basis the interaction vertex becomes
\begin{align}
  \hat  H'_{\rm int} = \sum_{\alpha,\beta \in \{e,f\}}\int_{\bs p}\sum_{\bs k \neq \bs 0} \alpha^\dagger_{\bs p+\bs k} \beta_{\bs p} \left [ g_{\alpha\beta}^{(+)} b^\dagger_{\bs k}+ g_{\alpha\beta}^{(-)} b_{-\bs k} \right ] ,
    \label{eq:1}
\end{align}
where $g_{\alpha\beta}^{(\pm)}$ (implicitly functions of $\bs p,\bs k$) are combinations of $u,v,\tilde u,\tilde v$, and $g$ (see supplement \cite{supp}). 

Due to the hybridization of the bands the effective mass changes sign at the point of maximum hybridization of trions and holes, see Fig.~\ref{fig:1} (b). As a consequence, the corresponding Hall resistivity which is controlled by the effective mass also changes sign when tuning the Fermi level into the regime of strong hybridization, which provides a striking signature of exciton condensation; see Fig.~\ref{fig:1} (c).

\textit{\textbf{Modeling exciton condensation --}} We consider a strictly 2D bosonic system with weak (repulsive) interaction strength $\sf u$, undergoing a BKT transition at a temperature $T_\star$. A mean-field treatment is nonetheless a reasonable first approximation, as the coherence length diverges exponentially with inverse temperature and hence will at some scale exceed the size of the sample. Below this temperature scale, quantum phase fluctuations  may be neglected in the superfluid regime. Related mean-field approaches were for example successfully used to capture experimentally measured spin-susceptibilities as well~\cite{qi2026}. 

Under this approximation, the total number of excitons, which is fixed experimentally, decomposes as $n_x = n_0 + n_{\rm q} + n_{\rm th}$, that are the condensed fraction and the quantum and classical fluctuations, respectively. Here, $n_{\rm q}=n_0 m_x {\sf u}/2\pi$. The thermal fluctuations 
are divergent in two dimensions (signaling the absence of genuine mean-field condensation). However, they can be regularized in a physically consistent way (see supplement \cite{supp}) such that the condensed fraction is linear in temperature, 
\begin{equation}
    n_0 =n_{x}\frac{1-T/T_\star}{1+{\sf u}m_x /2\pi} .
    \label{eq:cond}
\end{equation}
where the transition temperature $T_\star$ is proportional to the total exciton density \cite{supp,PhysRevB.37.4936, PhysRevLett.87.270402}.

\textit{\textbf{Kinetic theory --}} We study the transport properties of the itinerant charge carriers in the top layer, including the effect of interactions $H'_{\rm int}$, using a kinetic theory. We incorporate a transverse magnetic field $\bs B = \bs \nabla \times \bs A$, which affects holes and trions alike, but does not couple to the charge neutral excitons. Only the top layer is gated, hence transport is driven by a potential imbalance $V(x)=-E_x x$ affecting holes only. Technically, this means that the energy-momentum coordinates are replaced as $(\epsilon_h,\bs p) \rightarrow (\epsilon_{h}+{\sf e}V(x),\bs p - {\sf e}\bs A)$ for holes and $(\epsilon_{t},\bs p) \rightarrow (\epsilon_{t},\bs p - {\sf e}\bs A)$ for trions, prior to diagonalization. In a previous work Ref.~\cite{Zerba2025}, we showed that, in the absence of a magnetic field, a hole current induces drag transport of trions and excitons. Here, the band reconstruction modifies the picture: each quasiparticle species ($\hat e^\dagger,\hat f^\dagger$) couples to the electric field via a fraction $\xi_{e/f}\in[0,1]$ of the hole charge, $\xi_{e/f} = (1 \pm (\epsilon_h-\epsilon_t)/[(\epsilon_h-\epsilon_t)^2+4\sdelta^2]^{1/2})/2$, and thus has its own Hall ratio, that importantly will depend on the quasiparticle's effective mass. Similarly, both $\hat e^\dagger$ and $\hat f^\dagger$ contribute to the hole current, i.e., the charge current in the top layer, 
\begin{equation}
\label{eq:hole_current}
  \bs J_h = \sum_{\alpha \in \{e,f\}}\int_{\bs p} \xi_\alpha(\bs p) \langle \alpha_{\bs p}^\dagger \alpha_{\bs p} \rangle \,{\sf e}\frac{ \partial\epsilon_h}{\partial {\bs p} },
\end{equation}
where $\langle \cdot \rangle$ is the non-equilibrium average.
At $n_0=0$ where the bands do not hybridize, this current is carried solely by the holes, even when the trion population is put off-equilibrium by drag effects.

To determine the populations $\langle \alpha_{\bs p}^\dagger \alpha_{\bs p} \rangle$, we first derive a set of coupled Boltzmann equations for the three quasiparticle species $(\hat b^\dagger,\hat e^\dagger,\hat f^\dagger)$, where the collision integral is determined by $H'_{\rm int}$, Eq.~\eqref{eq:1}. We also add a constant relaxation rate $\tau$ that accounts for  disorder scattering, which we assume affects all species of particles in the same way; our results do not depend qualitatively on the value of $\tau$, which we set numerically to $\tau= 10 $ ps, motivated by transport experiments~\cite{Joe2024,Pack2024Jul,Guo2024}. We then numerically solve the system of Boltzmann equations to obtain the particle populations, and in turn the current.
From the linear response relation $E_x = \rho_{x\mu}^hJ_\mu$ we extract the longitudinal and Hall resistivities, $\rho_{xx}^h$ and $\rho_{xy}^h$, which we will now discuss. 

\begin{figure}
    \centering
    \includegraphics[width =\linewidth]{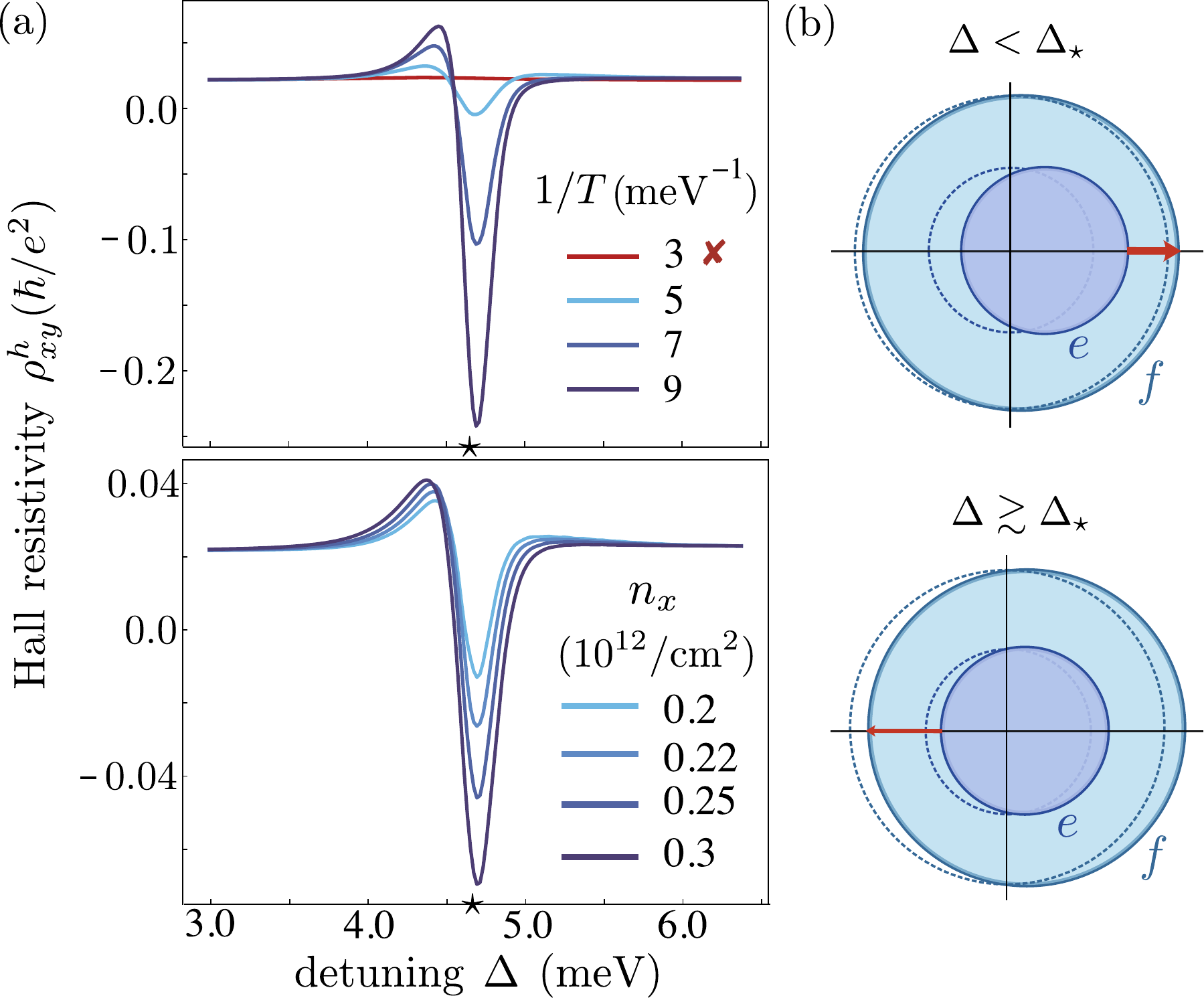}
    \caption{\textbf{Hall resistivity $\rho_{xy}^h$.} (a) Hall resistivity $\rho_{xy}^h$ as a function of detuning $\Delta$ for different values of temperature $T$ at $n_x=0.25 \,\cdot 10^{12}\;{\rm cm}^{-2}$ (top) and exciton density $n_x$ at $1/T = 6$ meV$^{-1}$ (bottom). The red curve corresponds to the absence of the condensate. (b) Physical picture for kinetics and scattering mechanisms. The momentum-space displacement of each Fermi sea $\alpha \in \{ e,f\}$ depends on its effective charge $\xi_{\alpha} \sf e$ and hence on $\Delta$. Off resonance (top) only one Fermi surface is significantly displaced, while near resonance (bottom) both are. This determines the strength of the interband scattering (represented by the width of the red arrow), 
    which in turn determines the asymmetry of $\rho_{xy}^h$ around the resonance.} 
    \label{fig:temp_nx_Hall}
\end{figure}

\textit{\textbf{Sign reversal of Hall response --}} 
A crucial consequence of the hole-trion hybridization is that at the avoided crossing, the effective masses of fermions in each branch, determined by the curvature of the dispersion relation as $m_\alpha^\star = (\partial_{\bs p}^2 \epsilon_{\alpha})^{-1}$, $\alpha \in \{e,f\}$, have opposite signs. In particular, close to resonance at $\Delta_\star \approx \tfrac 2 3 \mu_h$~\cite{Zerba2025}, where the Fermi level coincides with the avoided crossing, most of the hole current $\bs J_h$ is carried by
excitations whose mass changed sign. 
It can already be seen from the Drude picture for Hall resistivity, $\rho_{xy}=B\pi/{\sf e}m^\star\mu$, that this affects the sign of the Hall coefficient directly. 

To witness this, we evaluate the Hall resistivity $\rho^h_{xy}$ of holes within the kinetic theory as a function of detuning $\Delta$, Fig.~\ref{fig:1} (c), both with and without an exciton condensate (the latter is obtained by taking $n_0 \rightarrow 0$), at the same temperature $T$ and for the same exciton density $n_x$. For our numerical evaluations we use, unless specified otherwise, the physically motivated parameters $\mu_h=7$ meV, $n_x=0.25 \,\,10^{-12} \text{cm} ^{-2}$, ${\sf u}= 4$ meV, and $B= 10^{-1}$ T; the robustness of our results as a function of the magnetic field is analyzed in detail in the End Matter section A. 
The interaction constant $g \simeq \left ( {3\pi} / m_h |E_t^0| \right )^{1/2} {\sf t}$, where we take ${\sf t}= 0.3$ meV and $E_t^0 = - 10$ meV \cite{schwartz21, Nguyen2025, Qi2025drag, Kuhlenkamp21, Zerba2024, Zerba2025}. In the absence of an exciton condensate, the non-interacting quasiparticles are holes, trions, and excitons: the problem then reduces to the one we studied in Ref.~\cite{Zerba2025}, except now in a magnetic field. The resulting $\rho^h_{xy}$ curve shows a small peak near resonance in the full kinetic treatment, resulting from many-body interactions that are enhanced at resonance; by contrast, in the Drude approximation  $\rho^h_{xy}$ would be independent from relaxation rates and thus from the detuning $\Delta$. 
In the presence of a condensate, the strongly peaked structure of $\rho^h_{xy}$ as a function of detuning $\Delta$ shows a striking sign change of the Hall resistivity at resonance. This is a direct consequence of the band reconstruction. The effective \emph{hole} mass depends on the participation ratio of the $(e,f)$ particles to the hole current: Fig.~\ref{fig:1} (c) (inset) shows that holes have a positive effective mass at higher $\mu_h$ (or equivalently $\Delta \lesssim \Delta_\star$), and a negative effective mass at lower $\mu_h$ ($\Delta \gtrsim \Delta_\star$). This explains the sign-structure of the peaks as a function of $\Delta$. The sign reversal is a direct consequence of exciton condensation and constitutes our main result.

We will now analyze the features that are direct consequences of the band reconstruction by the boson condensation. Such features depend on $n_x$ and $T$ mostly which in turn modify $n_0$.
We find that the size of the peaks in $\rho_{xy}^h$ increases rapidly with $1/T$, see Fig.~\ref{fig:temp_nx_Hall} (a). This is primarily because the effect of thermal broadening is to average thermal transport over the two bands with opposite local curvatures, which softens the signatures of the hybridization; Fig.~\ref{fig:1} (c), inset. This thermal activation mechanism explains the rapid growth of the $\rho^h_{xy}$ peaks with $1/T$. 
Moreover, at low temperatures the condensed density $n_0$ and thus the direct hybridization gap $\sdelta$ is larger, which further reduces such thermal smearing effects and ensures clear peaked features. We note that a larger $\sdelta$ also reduces the band curvature (in absolute terms) near the avoided crossing, which in principle would have an opposite effect on $\rho_{xy}^h \propto m_{\star}^{-1}$, however, we find numerically that for the range of parameters considered, this effect is negligible.

The dependence of $\rho_{xy}^h$ on the total (fixed) exciton density $n_x$ is strongest on the large negative peak. 
At fixed intermediate temperature $T$, the thermal activation mechanism leads to a peak height $\sim e^{\sdelta/T}$ where $\sdelta = g\sqrt{n_0}\propto g\sqrt{n_x}$, as seen from  Eq.~\eqref{eq:cond}. Hence, the hight of the peak increases with  $n_x$, see Fig.~\ref{fig:temp_nx_Hall} (a). 

As a next step, we explore the effects of interactions of fermions with sound modes of the exciton condensate. 
A distinctive feature of $\rho_{xy}^h$ is that the negative peak is much larger than its positive counterpart. The imbalance between the two peaks is a combined effect of the peak positions not being symmetric with respect to $\Delta_\star$, and of scattering being less efficient at resonance than away from it resonance. At resonance,  the two fermion species couple identically to the electric field, and their Fermi distributions are displaced out of equilibrium by an equal amount ($\xi_e= \xi_f=1/2 $ and $\bs v_e = \bs v_f$). Off resonance, $\xi_f \gg \xi_e$ (or vice versa), so mostly only one Fermi distribution is displaced out of equilibrium. The minimal boson momentum required to scatter from one to the other is therefore smaller away from resonance, see Fig.~\ref{fig:temp_nx_Hall} (b), which in turn ensures more efficient scattering. As the current is given by a contribution from both species, Eq.~\eqref{eq:hole_current}, this effect results in an imbalance between the peaks of $\rho_{xy}^h$, and a much larger negative peak; see also section B of the End Matter. 
\begin{figure}
    \centering
    \includegraphics[width =\linewidth]{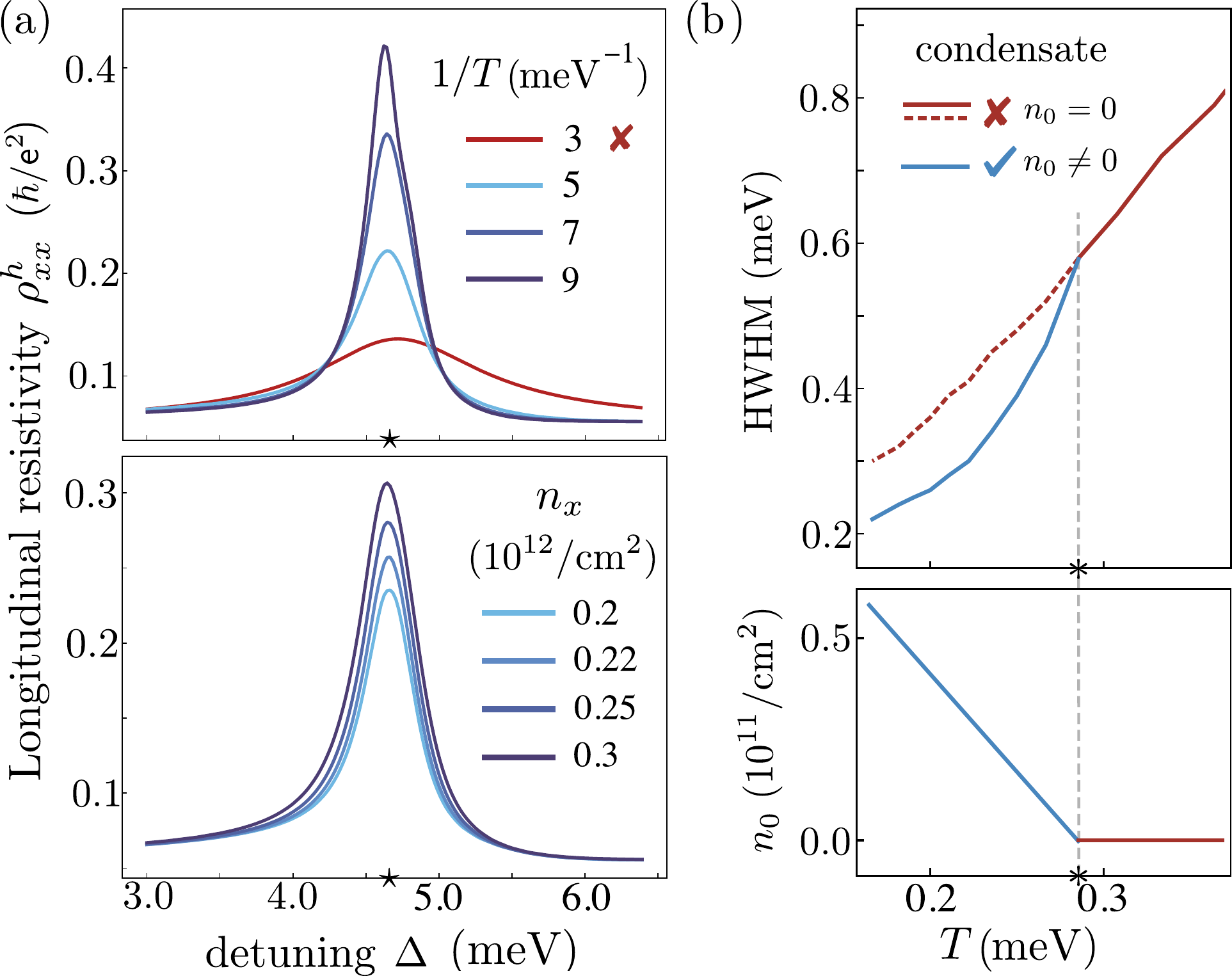}
    \caption{\textbf{Longitudinal resistivity.} (a) Longitudinal resistivity $\rho_{xx}^h$ as a function of detuning $\Delta$ for different values of temperature $T$ at $n_x=0.25 \,\cdot 10^{12}\;{\rm cm}^{-2}$ (top) and exciton density $n_x$ at $1/T = 6$ meV$^{-1}$ (bottom). The red curve corresponds to the absence of the condensate.  (b) Half Width at Half Maximum (HWHM) of the longitudinal resistivity (top) and condensate density $n_0$ (bottom). Upon condensation, the phase space available for scattering decreases compared to the non-condensed case (red dashed curve, obtained by taking $n_0 \rightarrow 0$ at fixed $n_x,T$), and the resonance region is narrowed. The cusp in HWHM thus detects exciton condensation. }
    \label{fig:temp_nx_longitudinal}
\end{figure}

Crucially, this line of reasoning concerns interband scattering and determines $\rho_{xy}^h$, which we argued peaks near \emph{but not at} resonance. Meanwhile,  $\rho_{xx}^h$ is mainly governed by intraband scattering, that always peaks at resonance.

\textit{\textbf{Modified scattering phase space from exciton condensation --}} Close to resonance, scattering with thermally excited bosons dominates transport regardless of condensation, so that $\rho_{xx}^h$ follows the universal $1/T$ behavior~\cite{Zerba2025}; see Fig.~\ref{fig:temp_nx_longitudinal} (a).
Similarly, the exciton density $n_x$ has significant effects on $\rho_{xx}^h$ only near resonance. There, increasing $n_x$ proportionally increases the population of bosonic excitations available for scattering, giving rise to an approximately linear increase of the peak value of $\rho_{xx}^h$ with $n_x$. 
Away from resonance, $\rho_{xx}^h$ depends weakly on $T$ or $n_x$, since there the boson-fermion scattering rate is much smaller than extrinsic disorder scattering $\tau$. However, strikingly upon condensation the available phase space for scattering abruptly reduces due to the appearance of linear sound modes of the exciton condensate; see Fig.~\ref{fig:temp_nx_longitudinal} (a) at $\Delta > \Delta_\star$ where the effect is strongest. 
For $\Delta \ll \Delta_\star$, the effect appears less pronounced. We find that this trend arises from finite momentum-space discretization in the numerical calculations, which becomes more relevant for large trion Fermi surfaces ($\Delta \ll \Delta_\star$).

As a consequence, the resonance peak becomes narrower when excitons condense. This is evidenced in the behavior of the Half Width Half Maximum (HWHM) of the longitudinal resistivity, see Fig.~\ref{fig:temp_nx_longitudinal} (b). 

Crucially, this reduction of phase space available for scattering induced by the change in bosonic dispersion upon condensation, is a general mechanism that can be used to detect exciton condensates also in simpler heterostructures than the one discussed here~\cite{Qi2025electrically,Nguyen2025,qi2026}. Even when interactions between excitons and charge carriers cannot be tuned (corresponding to a fixed $\Delta$), condensation of excitons will still induce a cusp in the longitudinal resistivity when reducing temperature below the effective condensation temperature.

\textit{\textbf{Conclusion \& Outlook --}} We have shown that charge transport provides a powerful probe of exciton condensation in semiconductor heterostructures. 
We have identified two signatures: (i) a reduction of the resistivity due to the suppression of carrier scattering by the condensate, and (ii) in the presence of a tunable solid-state Feshbach resonance, a condensate-induced reconstruction of the quasiparticle bands, producing a sign reversal of the Hall resistivity. Both are striking experimental signatures of exciton condensation in charge transport.

For future work it will be interesting to explore the spin-valley structure of excitons~\cite{qi2026} and the role of Berry curvature~\cite{Xiao2012, Xu2014}, which may give rise to additional transport phenomena. More broadly, the experimentally accessible, exceptionally low effective temperatures of the doped carriers, $T\lesssim0.01\,T_{\rm F}$~\cite{Qi2025electrically, Nguyen2025}, render these systems a promising platform for exploring strongly interacting Bose-Fermi mixtures under extreme conditions. There, interaction-mediated superconductivity~\cite{Zerba2024, Crepel2023, Milczewski2024, Kumar2025} and competing nematic or density-wave instabilities could emerge, opening exciting opportunities for both theoretical and experimental exploration.

\textit{\textbf{Acknowledgments --}} We  acknowledge insightful discussions and prior
collaborations with A. Imamo\u{g}lu and C. Kuhlenkamp. We acknowledge support from the Deutsche Forschungsgemeinschaft (DFG, German Research Foundation) under Germany’s Excellence Strategy–EXC–2111–390814868, TRR 360 – 492547816 and DFG grants No. KN1254/1-2, KN1254/2-1, the European Union (grant agreement No 101169765), as well as the Munich Quantum Valley, which is supported by the Bavarian state government with funds from the Hightech Agenda Bayern Plus.

\textit{\textbf{Data availability --}} Data and codes are available upon reasonable request on Zenodo~\cite{zenodo}.

\bibliography{library}

\appendix

\section*{End Matter}
\setcounter{equation}{0}
\renewcommand{\theequation}{EM~\arabic{equation}}

\subsection{A. Magnetoresistive effects }\label{em_mag}

\begin{figure}
    \centering
    \includegraphics[width=\linewidth]{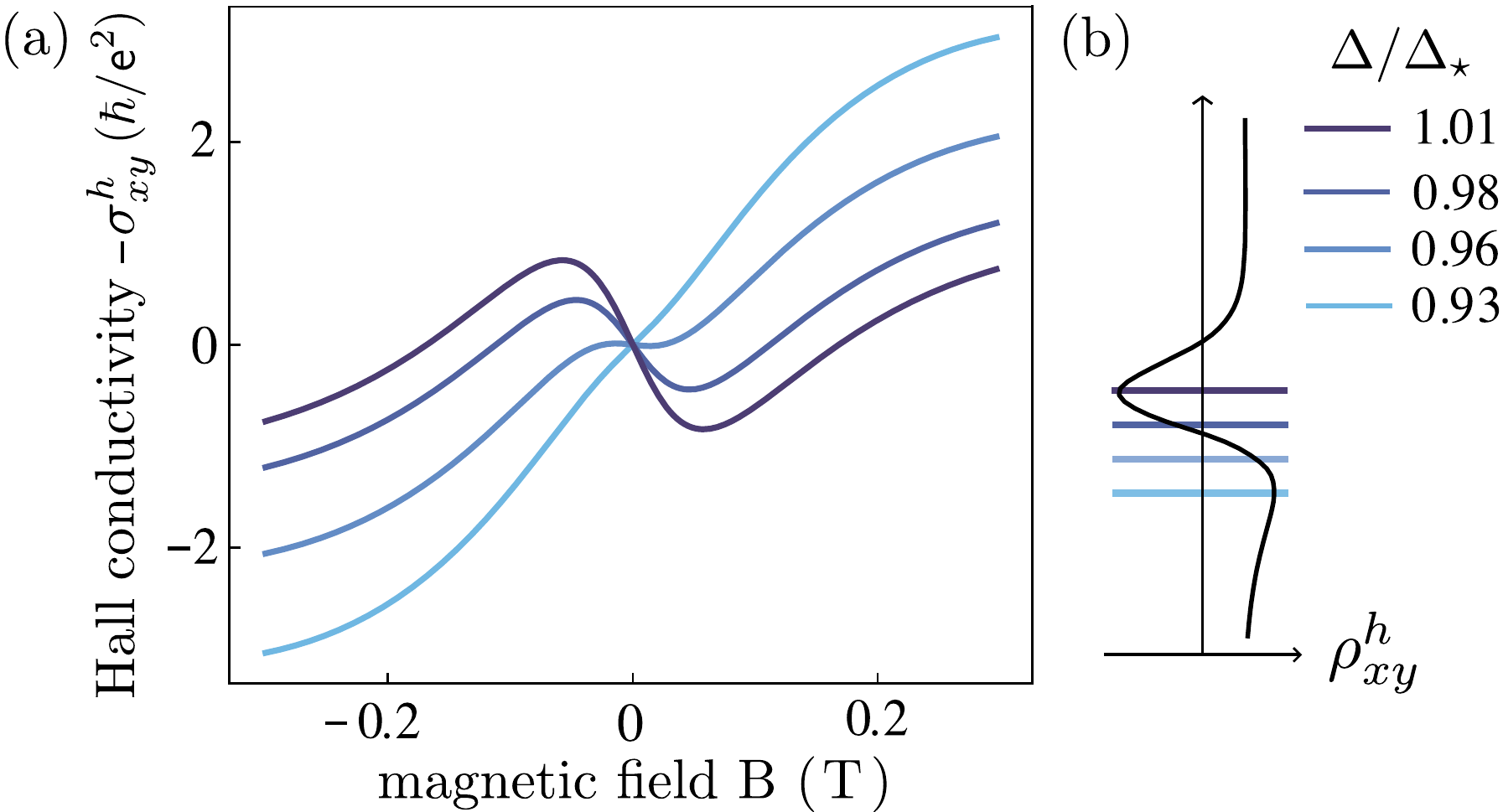}
    \caption{\textbf{Magnetic field dependence of the Hall conductivity $\sigma_{xy}^h$ near resonance}. (a) Negative Hall conductivity $-\sigma_{xy}^h$ of the holes in the presence of an exciton condensate as a function of the magnetic field $B$ for several values of $\Delta$ near resonance. At large $B$, the negative slope (a hallmark of condensation)  disappears, as a consequence of strong magnetoresistive effects. (b) Corresponding $\rho_{xy}^h$ at fixed $B=0.1$ T.}
    \label{fig:magnetic_field}
\end{figure}

We evaluate the Hall conductivity $\sigma_{xy}^h$ as a function of the magnetic field $B$; see
Fig.~\ref{fig:magnetic_field}. We find that the negative peak near resonance, signaling exciton condensation, is robust over a range of magnetic field values, but eventually disappears at large $B$. This arises from the interplay of strong magnetoresistive effects 
that become relevant when increasing the magnetic field.
The behavior of the Hall conductivity can be understood as arising from two fluids of fermions $\alpha \in \{e,f\}$, carrying different charges $\xi_{\alpha} {\sf e}$ and whose conductivities are additive. The total Hall conductivity is then
\begin{equation}
    \sigma_{xy}^h = \sum_{\alpha}\frac{\xi_\alpha{\sf e}B/m_\alpha}{(1/\tau_\alpha)^2 + ({\sf e}B/m_\alpha)^2},
\end{equation}
with $\tau_\alpha$ a species-dependent relaxation rate resulting from interactions and a constant background relaxation rate.
At small values of the magnetic field, $1/\tau_\alpha \gg {\sf e}B/m_\alpha$ for both $\alpha \in \{e, f\}$, so the Hall conductivity is linear in $B$,
\begin{equation}\label{tfr}
 \sigma_{xy}^h \approx \sum_{\alpha}\frac{\xi_\alpha{\sf e}B/m_\alpha}{(1/\tau_\alpha)^2} .
\end{equation}
The sign of the slope, at a given $\Delta$, is determined by the sign of ${\sf e}/m_\alpha$ where $\alpha$ is the lightest species at this $\Delta$ (that is, $f$ at the negative peak $\Delta_-$, and $e$ at the positive peak $\Delta_+$). This ensures the slope to be negative at $\Delta_-$ since $m_f(\Delta_-)<0$, and positive at $\Delta_+$ since $m_e(\Delta_+)>0$.

For large magnetic fields, $1/\tau_\alpha \ll {\sf e}B/m_\alpha$ for both $\alpha$, so the Hall conductivity is ultimately proportinoal to $1/B$,
\begin{equation}
\label{eq:sigmaxy_twofluid}
     \sigma_{xy}^h \approx \sum_{\alpha}\frac{\xi_\alpha}{{\sf e}B/m_\alpha} .
\end{equation}
Its sign at a given $\Delta$ is determined by the sign of ${\sf e}/m_\alpha$ where $\alpha$ is the \emph{heaviest} species at this $\Delta$ (that is, $e$ at the negative peak $\Delta_-$, and $f$ at the positive peak $\Delta_+$). Because $m_e(\Delta_-)>0$, this explains the sign change observed in Fig.~\ref{fig:magnetic_field}, i.e., the $-\sigma_{xy}^h(\Delta_-)$ initially negative at small fields becomes positive at larger fields.
Meanwhile, no such sign change is seen for those values of $\Delta$ where $-\sigma_{xy}^h$ is positive at small fields, as $m_f(\Delta_+)>0$.

\section{B. Asymmetry of Hall response}\label{em_np}

We provide an explanation for the asymmetry between the positive and negative peaks of $\rho_{xy}^h$ shown in Fig.~\ref{fig:temp_nx_Hall}. We emphasize that this is an interaction-driven effect. Indeed, in the absence of interactions, we find numerically that $\rho_{xy}^h$ displays two peaks with equal and opposite heights. Here, we explain the strong enhancement of the negative peak, as a result of a shift of the resonance induced by interactions and of phase-space constraints.

The behavior of the transverse resistivity can be understood as arising from the same two-fluid, Drude-like picture as above. The total Hall conductivity is given by Eq.~\eqref{eq:sigmaxy_twofluid} and the longitudinal conductivity is
\begin{align}
\label{eq:starting_pt_xy}
    \sigma_{xx} &= \sum_{\alpha}\frac{1/\tau_\alpha}{(1/\tau_\alpha)^2 + ({\sf e}B/m_\alpha)^2}. 
\end{align}

From this the Hall resistivity $\rho_{xy} \approx - \sigma_{xy}^h/(\sigma_{xx}^h)^2$ is obtained. 
In the above expressions, all species-dependent quantities are evaluated at their corresponding Fermi level, and are thus functions of $\Delta$. We note that, close to resonance, the magnetoresistance term in the denominator is negligible, $1/\tau_\alpha \gg {\sf e}B\xi_\alpha /m_\alpha$,  
so we approximate and find
\begin{align}
\label{tfr}
\rho_{xy} \approx - \sum_{\alpha} \frac{\xi_\alpha{\sf e}B/m_\alpha}{(1 + \tau_{\underline{\alpha}}/\tau_\alpha)^2} .
\end{align}
where $\underline{\alpha}$ indicates the complementary component of $\alpha$, namely $\alpha = e \Rightarrow {\underline \alpha} = f$ and conversely.

That the two peaks of $\rho_{xy}$ occur at different values of $\Delta$ reflects the fact that the maximal inverse effective band mass $1/m_\alpha$ is reached at different values of $\Delta$ for the two bands: one has $|m_f| \gg |m_e|$ at the positive peak, and $|m_e| \gg |m_f|$ at the negative peak. Thus, in Eq.~\eqref{tfr} it is sufficient to keep the term $\alpha=e$ for the positive peak, and $\alpha=f$ for the negative peak.
Because the extremal band curvatures are approximately symmetric, the ratio between the peak heights is determined by the ratio of the characteristic scattering times of the two species, $\tau_f/\tau_e$, evaluated at the values of detuning $\Delta_\pm$ corresponding to the two peaks of $\rho_{xy}$. The ratio between peak heights, 
\begin{equation}
    \frac{|\rho_{xy}(\Delta_-)|}{|\rho_{xy}(\Delta_+)|}\; \propto\; \frac {(1 + \tau_f/\tau_e)^2\big |_{\Delta_+} }{(1 + \tau_e/\tau_f)^2\big |_{\Delta_-}} ,
\end{equation}
would exhibit an enhancement of the negative peak provided $(\tau_f/\tau_e)|_{\Delta_+} >  (\tau_e/\tau_f)|_{\Delta_-} $.

The scattering rates contain contributions from intra-band scattering, inter-band scattering, and extrinsic scattering of rate $1/\tau$: 
\begin{equation}
    1/\tau_\alpha = 1/\tau + \mathcal I_{\alpha \alpha} + \mathcal I_{\alpha \underline{\alpha}} ,
\end{equation}
where $\mathcal I_{\alpha\beta}$ stand for the collision kernels.
The conservation laws during the scattering process ensure $\mathcal I_{ef}=-\mathcal I_{fe}$ at a given $\Delta$. Besides, since the dominant inter-band scattering process must relax the out-of-equilibrium distribution of the lightest species (that is $e$ at $\Delta_+$ and $f$ at $\Delta_-$), the corresponding collision terms are positive: $\mathcal I_{ef}(\Delta_+)>0$ and $\mathcal I_{fe}(\Delta_-)>0$. 
The intra-band collision integrals $\mathcal I_{\alpha\alpha}$ are proportional to the particle density of states whose behavior at resonance are symmetric, therefore $\mathcal I_{ff}(\Delta_\pm) \approx \mathcal I_{ee}(\Delta_\mp)$. 
In summary, the above condition for an enhanced negative peak can be recast as
\begin{equation}
\label{eq:condition}
    \frac{1/\tau + \mathcal I_{ee}^{(+)} + \mathcal I_{ef}^{(+)}}{1/\tau + \mathcal I_{ff}^{(+)} - \mathcal I_{ef}^{(+)}} 
    > 
    \frac{1/\tau + \mathcal I_{ee}^{(+)} +\mathcal I_{fe}^{(-)}}{1/\tau + \mathcal I_{ff}^{(+)} - \mathcal I_{fe}^{(-)}},
\end{equation}
where we use the shorthand $\mathcal I_{\alpha\beta}^{(\pm)} = \mathcal I_{\alpha\beta}(\Delta_\pm)$. In particular, Eq.~\eqref{eq:condition} is equivalent to $\mathcal I_{ef}^{(+)} > \mathcal I_{fe}^{(-)}$, as both quantities are positive. 

Upon lowering the temperature, the peak in the longitudinal resistivity $\rho_{xx}^h$ is shifting to lower $\Delta$ and so does the envelop of $\rho_{xy}^h$. This ensures that  $\Delta_- \approx \Delta_\star$ while $\Delta_+$ is pushed further away from resonance as an effect of interactions; see Fig.~\ref{fig:temp_nx_Hall} (a). 
The collision kernels $\mathcal I_{ef}$, $\mathcal I_{fe}$ describe scattering between the two different Fermi surfaces, and depend sensitively on the position $\Delta_\pm$ of the peaks with respect to the bare resonance $\Delta_\star$. 
Since the two Fermi surfaces are displaced by a same amount near resonance $\Delta_\star$, but by different amounts away from resonance, the bosons involved in the dominant scattering process leading to $\mathcal I_{ef}^{(+)}$ carry smaller momenta than those involved in $\mathcal I_{ef}^{(-)}$; see Fig.~\ref{fig:temp_nx_Hall} (b). Thus the relation $\mathcal I_{ef}^{(+)} > \mathcal I_{fe}^{(-)}$ is satisfied, which explains the asymmetry between the two peaks, with a much larger negative than positive peak.

\end{document}


\newcommand{\ourtitle}{Supplemental Material to Detecting Exciton Condensation through Charge Transport in Semiconductor Heterostructures}
\title{\ourtitle}

\newcommand{\TUM}{\affiliation{Technical University of Munich, TUM School of Natural Sciences, Physics Department, 85748 Garching, Germany}}
\newcommand{\MCQST}{\affiliation{Munich Center for Quantum Science and Technology (MCQST), Schellingstr. 4, 80799 M{\"u}nchen, Germany}}
\newcommand{\Harvard}
{\affiliation{Department of Physics, Harvard University, Cambridge, Massachusetts 02138, USA}}

\author{Caterina Zerba} 
\TUM 
\MCQST 

\author{L\'eo Mangeolle} 
\TUM 
\MCQST

\author{Michael Knap} 
\TUM 
\MCQST

\maketitle

\setcounter{equation}{0}
\renewcommand{\theequation}{S~\arabic{equation}}

\section{Model for the quasi-condensate}
\label{sec:details-condensate-model}

The number of excitations above the condensate is 
\begin{align}
    n_x - n_0 &= \sum_{\mb k \neq \bs 0} \langle x_{\mb k}^\dagger x_{\mb k} \rangle 
    = n_{\rm q} + n_{\rm th} ,
\end{align}
where 
\begin{align}
n_{\rm q}  = \sum_{\mb k \neq \bs 0}   v _{\mb k} ^2 
&= \int \frac{\text d^2 \bold k }{(2\pi)^2}\Big (\frac{E^x_{\bold k }+{\sf u}n_0}{\sqrt{E^x_{\bold k }(E^x_{\bold k }+2 {\sf u} n_0)}}-1\Big )= ({\sf u} m_x n_0/\pi) \int_0^\infty \text dy y\Big(\frac{y^2 +1 }{\sqrt{(y^2 (y ^2 +2))}}-1\Big) =
n_0  {\sf u} m_x/ 2\pi
\end{align}
and
\begin{align}
n_{\rm th} = \sum_{\bold k \neq 0} (  v_{\bold k } ^2+  u _{\bold k } ^2)\langle b_{\bold k }^\dagger b_{\bold k }\rangle
&= \int \frac{\text d^2 \bold k }{(2\pi)^2} \bigg(\frac{E^x_{\bold k }+{\sf u} n_0 }{\sqrt{E^x_{\bold k }(E^x_{\bold k }+2 {\sf u} n_0 )}}\bigg)n_{\rm B}(\omega_{\bold k}) 
= \int _0 ^\infty \text dy y \frac{y ^2  +1}{\sqrt{y^2(y^2 +2)}} \frac{({\sf u}m_xn_0/\pi)}{e^{\beta {\sf u}n_0 \sqrt{y^2(y^2+2) }}-1} ,
\end{align}
 where in both cases the change of variables to $y = k/\sqrt{2{\sf u}n_0m_x}$ is performed.

The latter integral has logarithmic divergence at $y\rightarrow 0$,
reflecting the divergence of thermal fluctuations in $d=2$, equivalently the Hohenberg-Mermin-Wagner theorem preventing the formation of a genuine condensate. 

To model the quasi-condensate with finite $n_0(T)$, it is thus necessary to introduce a physical infrared cutoff $\Lambda^{-1}(T)$ in the above integral. The total contribution to the integral by $y \in [T/{\sf u}n_0,+\infty [$ is negligible, so $T/{\sf u}n_0$ can be used as an upper bound. Assuming low temperatures $T/{\sf u}n_0 \ll 1$, one is left with
\begin{equation}
 n_{\rm th} \approx (m_x {\sf u}n_0/\pi) \int_{\Lambda^{-1}(T)}^{T/{\sf u}n_0} 
 \text dy y \frac 1 {2 {\sf u} n_0 y^2/T}  = (m_x T/2\pi) \ln(T \Lambda/ {\sf u}n_0) .
\end{equation}

The cutoff may now be fixed phenomenologically to ensure the correct dependence $n_0(T)$. 
Here we assume a dependence $n_0 = \gamma(T_\star - T)$. 
This unambiguously fixes the cutoff
\begin{align}
    \Lambda(T) = \gamma {\sf u} (T_\star/T - 1) \exp \left \{ (2\pi/m_xT)\left [ n_x - \gamma(T_\star - T)(1+{\sf u}m_x/2\pi)\right ] \right \} .
\end{align}
In particular, the constraints $n_{\rm th}(T=0)=0$ and $n_0(T=T_\star)=0$ imply the result quoted in the main text.
For the transition temperature, we use the formula $T_\star = \frac{2\pi \hbar^2 n_x}{m_x k_{\rm B}}\left [ \ln(K/4\pi) + \ln(\ln(1/n_x a))\right ]^{-1}$ where $K \approx 380$ is a constant determined numerically, and $a$ is the scattering length, related to he interaction strength $\sf u$ as $a={\sf u} m_x/4\pi\hbar^2$ \cite{PhysRevB.37.4936, PhysRevLett.87.270402}.

\section{Band reconstruction}
\label{sec:details-bogoliubov-hybrid}

The model we consider contains three species of particles: holes $h^\dagger_{\bs p}$, excitons $x^\dagger_{\bs k}$, and trions (= molecules) $m^\dagger_{\bs p}$. Their dynamics and interactions are described by $H = H_{0,\rm f} + H_{\rm b} + H_{\rm int}$, where the fermionic free Hamiltonian is
\begin{equation}
  H_{0,\rm f} = \int_{\bs p}\epsilon_{h}(\bs p) \, h^\dagger_{\bs p}h_{\bs p} \;+\;
    \int_{\bs p}\epsilon_{t}(\bs p) \, m^\dagger_{\bs p}m_{\bs p} 
\end{equation}
and the bosonic sector's hamiltonian is
\begin{equation}
 H_{\rm b} =   \int_{\bs k}\epsilon_{x}(\bs k) \, x^\dagger_{\bs k} x_{\bs k} 
    \;+\; {\sf u} \int_{\bs k,\bs k',\bs q} x^{\dagger}_{\bs{k'+q}} x^{\dagger}_{\bs{k-q}} x_{\bs k'}x_{\bs k} .
\end{equation}

 We model the formation of a quasi-condensate at low temperatures in the Bogoliubov approximation, assuming a finite $\langle x_{\bs k=\bs 0} \rangle=\sqrt n_0$. 
 Bosonic excitations $b^\dagger_{\bs k}$ above the condensate have dispersion 
 $\omega({\bold k })=\sqrt{(E_{\bold k}^{x})^2 - ({\sf u}n_0)^2}$, where $E_{\bold k}^{x} = \epsilon_x({\bold k}) + 2 {\sf u} n_0$ and we recall $\epsilon_x(\bs k)= k^2/2m_x - \mu_x$. In the following, at the mean-field level we take $\mu_x = {\sf u}n_0$. 
 These bosonic excitations are related to the exciton creation-annihilation operators by the transformation 
 $  x_{\bs k} = \tilde u_{\bs k} b_{\bs k} - \tilde v_{\bs k} b^\dagger_{-\bs k}$ (at $\bs k \neq \bs 0$),
with real parameters $\tilde u_{\bs k}, \tilde v_{\bs k}$ satisfying $\tilde u_{\bs k}^2 - \tilde v_{\bs k}^2 =1$, 
 defined as
 \begin{equation}
    \tilde u_{\bold k}=\frac 1 {\sqrt 2} \big [ (E^x_{\bold k})^2/((E^x_{\bold k})^2-({\sf u}n_0)^2)+1 \big ]^{1/2},\qquad \tilde v_{\bold k}= \frac 1 {\sqrt 2} \big [ (E^x_{\bold k})^2/((E^x_{\bold k})^2-({\sf u}n_0)^2)-1 \big ]^{1/2} .
 \end{equation}

The presence of a condensate also entails hybridization of the fermionic bands, namely
\begin{equation}
    H_{0,\rm f} + H_{\rm hyb} 
    = \int_{\bs p} \lambda_{e}(\bs p)\, e^\dagger_{\bs p} e_{\bs p}
    + \int_{\bs p} \lambda_{f}(\bs p)\, f^\dagger_{\bs p} f_{\bs p} ,
\end{equation}
where the redefined fermionic excitations have dispersion
\begin{align}
\lambda_{e}(\bs p) &= \frac 1 2 [ \epsilon_{h}(\bs p) + \epsilon_{t}(\bs p) ]
+ \sqrt{[\epsilon_{h}(\bs p) - \epsilon_{t}(\bs p)]^2/4 + g^2n_0 } ,\\
\lambda_{f}(\bs p) &= \frac 1 2 [ \epsilon_{h}(\bs p) + \epsilon_{t}(\bs p) ]
- \sqrt{[\epsilon_{h}(\bs p) - \epsilon_{t}(\bs p)]^2/4 + g^2n_0 } .
\end{align}
Their annihilation operators are related to those of holes and trions by a unitary transformation,
 \begin{align}
  \begin{cases}
    e_{\bs p} &= v^*_{\bs p} m_{\bs p}+u^*_{\bs p} h_{\bs p} \\
     f_{\bs p} &= u_{\bs p} m_{\bs p}-v_{\bs p} h_{\bs p} 
  \end{cases}\qquad \Leftrightarrow \qquad
  \begin{cases}
      m_{\bs p} &= u^*_{\bs p} f_{\bs p}+v_{\bs p} e_{\bs p}\\
h_{\bs p} &= -v^*_{\bs p} f_{\bs p}+u_{\bs p} e_{\bs p}
  \end{cases}
 \end{align}
 with coefficients $u_{\bs p}$ and $v_{\bs p}$ satisfying $|u_{\bs p}|^2+|v_{\bs p}|^2=1$ and defined as
\begin{equation}
    u_{\bs p}= \left [ 1 + g^2n_0/\varsigma(\bs p)^2 \right ]^{-1/2} ,
 \qquad v_{\bs p}=-\text{sign}[\varsigma(\bs p)] \, \left [ 1 + \varsigma(\bs p)^2/g^2n_0 \right ]^{-1/2}
\end{equation}
where we used the shorthand $\varsigma(\bs p) = \epsilon_t(\bs p) - \lambda_e(\bs p) 
  =   \lambda_f(\bs p) -  \epsilon_h(\bs p)$.

The interaction term then takes the form $H'_{\rm int}$ given in the main text, with
\begin{subequations}
\begin{align}
    g_{ff}^{(+)}(\bs p;\bs k) &= (u_{\bs{p+k}} v_{\bs p}^* \tilde v_{\bs k}^* - v_{\bs{p+k}} u_{\bs p}^* \tilde u_{-\bs k})g ,\\
    g_{ff}^{(-)}(\bs p;\bs k) &= (- u_{\bs{p+k}} v_{\bs p}^* \tilde u_{\bs k}^* + v_{\bs{p+k}} u_{\bs p}^* \tilde v_{-\bs k} )g,\\
    g_{fe}^{(+)}(\bs p;\bs k) &= (- u_{\bs{p+k}} u_{\bs p} \tilde v_{\bs k}^* - v_{\bs{p+k}} v_{\bs p} \tilde u_{-\bs k} )g,\\
    g_{fe}^{(-)}(\bs p;\bs k) &=  (u_{\bs{p+k}} u_{\bs p} \tilde u_{\bs k}^* + v_{\bs{p+k}} v_{\bs p} \tilde v_{-\bs k})g ,\\
    g_{ef}^{(+)}(\bs p;\bs k) &= (v_{\bs{p+k}}^* v_{\bs p}^* \tilde v_{\bs k}^* + u_{\bs{p+k}}^* u_{\bs p}^* \tilde u_{-\bs k} )g,\\
    g_{ef}^{(-)}(\bs p;\bs k) &= (- v_{\bs{p+k}}^* v_{\bs p}^* \tilde u_{\bs k}^* - u_{\bs{p+k}}^* u_{\bs p}^* \tilde v_{-\bs k} )g,\\
    g_{ee}^{(+)}(\bs p;\bs k) &= (-v_{\bs{p+k}}^* u_{\bs p}\tilde v_{\bs k}^* + u_{\bs{p+k}}^* v_{\bs p} \tilde u_{-\bs k})g ,\\
    g_{ee}^{(-)}(\bs p;\bs k) &= ( v_{\bs{p+k}}^* u_{\bs p} \tilde u_{\bs k}^* - u_{\bs{p+k}}^* v_{\bs p} \tilde v_{-\bs k})g .
\end{align}
\end{subequations}

One can check explicitly the symmetry $g_{\alpha\beta}^{(\pm)}(\bs p+\bs k;-\bs k) = \big [ g_{\beta \alpha}^{(\mp)}(\bs p;\bs k) \big ]^*$, which is also clear by definition from the hermiticity of $H'_{\rm int}$ defined in the main text. Thus, in the following we will express all quantities in terms of $g^{(-)}$ only.

\section{Kinetics}

First we discuss how the hole current operator can be written in terms of the populations of $e,f$ fermions. As explained in the main text, holes and trions couple to independent vector potentials $\bs A_h,\bs A_t$ and the hole current is defined from the derivative of the partition function $Z$ with respect to the former, $\bs J_h = T \partial \ln Z/\partial \bs A_h$. 
Equivalently,
\begin{equation}
\label{eq:Jh}
    \bs J_h = T \int_{\bs p}\frac{\delta \ln Z}{\delta \epsilon_h(\bs p)} 
    \frac{\partial \epsilon_h}{\partial \bs p} (-{\sf e})
    = -{\sf e}T \int_{\bs p}\sum_{\alpha \in \{e,f\}} \frac{\delta \ln Z}{\delta \lambda_\alpha(\bs p)} \frac{\partial \lambda_\alpha}{\partial \epsilon_h}
    \frac{\partial \epsilon_h}{\partial \bs p} ,
\end{equation}
which identifying $\delta \ln Z/\delta \lambda_\alpha(\bs p) = - (1/T)\langle \alpha^\dagger_{\bs p}\alpha_{\bs p} \rangle$ and $\partial \lambda_\alpha / \partial \epsilon_h = \xi_\alpha$ defined in the main text, leads to the main text's Eq.~(3). Here $\langle \alpha^\dagger_{\bs p}\alpha_{\bs p} \rangle$ denotes the non-equilibrium distribution of the $\alpha$ species.

Now we turn to the kinetic term of the Boltzmann equations, in the presence of electric and magnetic fields. In the basis $(e,f)$ of proper fermionic excitations, the time evolution of the on-shell projected density function $F_\alpha(\bs p, \bs X,t)$ is given by Boltzmann's equation, and the same is true for bosonic excitations $(b)$:
\begin{subequations}
\label{eqs:boltzmann}
\begin{align}
\big ( \partial_t + \bs v_\alpha(\bs p) \cdot  \partial_{\bs X}  + \bs f_\alpha(\bs p)\cdot  \partial_{\bs p} \big )\, F_\alpha &= I_\alpha \big [F_b,F_e,F_f \big ] \qquad \text{at}\, (\bs p,\bs X,t), \forall \alpha \in \{e,f\},\\
\big ( \partial_t + \bs v_b(\bs k) \cdot  \partial_{\bs X} \big )\, F_b &= I_b \big [F_b,F_e,F_f \big ] \qquad \text{at}\, (\bs k,\bs X,t),
\end{align}
\end{subequations}
where the square brackets indicate that different particle momenta are involved within the collision integral. 

The velocities are defined as $\bs v_\alpha = \partial_{\bs p} \lambda_\alpha$ and $\bs v_b = \partial_{\bs k}\omega$.
The Lorentz force acting on the fermions is 
\begin{equation}
   \bs f_\alpha(\bs p) = - \partial_{\bs X} \lambda_\alpha + {\sf e} \,\bs v_\alpha \times \bs B 
   = {\sf e}\left ( \xi_\alpha \bs E + \bs v_\alpha(\bs p) \times \bs B \right ) ,
\end{equation}
where we recall that $\xi_\alpha = \partial \lambda_\alpha/\partial \epsilon_h$ stems from the holes being affected by the longitudinal electric field $\bs E$ solely in the top layer, whence a fraction $\xi_\alpha \in [0,1]$ of their charge being inherited by each of the $(e,f)$ species. Meanwhile we assume the magnetic field $\bs B$ is felt by holes and trions alike, hence by both $(e,f)$ fermions.

We now expand the distribution function away from equilibrium, $F_{\alpha}(\bs p) = F_{\alpha}^{\rm eq}(\bs p)-2{\sf g}_{\alpha}(\bs p)$ for fermions and $F_{b}(\bs k)=F_{b}^{\rm eq}(\bs k)+2{\sf g}_{b}(\bs k)$ for bosons. Here equilibrium distributions are $F_{\alpha}^{\rm eq}(\bs p) = 1-2n_{\rm F}(\lambda_{\alpha}(\bs p))$ for fermions and $F_{b}^{\rm eq}(\bs k) = 1+2n_{\rm B}(\omega_{\bs k})$ for bosons.
The expansion of the left-hand side of Boltzmann's equation involves, in particular,
\begin{equation}
    {\sf e} (\bs v_\alpha \times \bs B)\cdot \partial_{\bs p} n_{\rm F}(\lambda_{\alpha})
    =   {\sf e} (\bs v_\alpha \times \bs B)\cdot \bs v_\alpha\, n_{\rm F}'(\lambda_{\alpha}) = 0 ,
\end{equation}
because of the vanishing triple product, and we note $ n_{\rm F}' = \partial_\lambda  n_{\rm F}$. 
Because ${\sf g}=O(\bs E)$, the term $\bs E \cdot \partial_{\bs p}{\sf g}=O(\bs E^2)$ may be neglected and the fermions' Boltzmann equations will thus take the form
\begin{equation}
\label{eq:lhs-in-field}
\partial_t {\sf g}_{\alpha} + \xi_\alpha {\sf e}(\bs E \cdot \bs v_\alpha) n'_{\rm F}(\lambda_\alpha) 
- {\sf e}(\bs B \times \bs v_{\alpha})\cdot \partial_{\bs p} {\sf g}_\alpha 
= - (1/2) I_{\alpha} ,
\end{equation}
and we will now proceed to computing $I_{\alpha} $.

\section{Derivation of the collision integrals}

We obtain the collision integrals using the standard real-time formalism for non-equilibrium many-body physics. 
We define fermionic $c^\pm$ and bosonic $b^\pm$ fields along a forward $(+)$ and backward $(-)$ time contour, and their conjugates $\bar c^\pm,\bar b^\pm$. We then perform the Keldysh rotations
\begin{subequations}
\begin{align}
  \label{eq:2}
  b_{\rm cl/q} &=  \left [ b^+ \pm^{\rm cl}_{\rm q} b^- \right ]/\sqrt 2 ,\\
 \bar b_{\rm cl/q} &=  \left [ \bar b^+ \pm^{\rm cl}_{\rm q} \bar b^- \right ]/\sqrt 2 ,\\
 c_{\rm r/a}  &= \left [ c^+ \pm^{\rm r}_{\rm a} c^- \right ]/\sqrt 2 ,\\
 \bar c_{\rm r/a}  &= \left [ \bar c^+ \pm^{\rm r}_{\rm a} \bar c^- \right ]/\sqrt 2 .
\end{align}
\end{subequations}

The interacting action in terms of these
classical/quantum and retarded/advanced fields reads
\begin{align}
     \label{eq:7}
     S_{\rm int} &= 
  - \int_{-\infty}^{+\infty} \text dt \frac 1 {\sqrt{2}} \sum_{\alpha,\beta \in \{c,d\}} \int_{\bs p}\int_{\bs k} g_{\alpha\beta}^{(-)}
                \left \lgroup
                b_{\bs k}^{\rm cl} \bar \alpha_{\bs p + \bs k}^{\rm a} \beta_{\bs p}^{\rm a}
                + b_{\bs k}^{\rm cl} \bar \alpha_{\bs p + \bs k}^{\rm r}  \beta_{\bs p}^{\rm r}
                + b_{\bs k}^{\rm q} \bar \alpha_{\bs p + \bs k}^{\rm a} \beta_{\bs p}^{\rm r} 
                  +  b_{\bs k}^{\rm q} \bar \alpha_{\bs p + \bs k}^{\rm r} \beta_{\bs p}^{\rm a}  \right \rgroup +\rm h.c ,
\end{align}
where ``+ h.c.'' means conjugating $b \mapsto \bar b$ and replacing $g_{\alpha\beta}^{(-)} \mapsto g_{\alpha\beta}^{(+)}$.
An interpretation of these interaction terms, in terms of diagrams, was provided in Ref.~\cite{Zerba2025}. Now, in Eq.~\eqref{eq:7}, the crucial difference is that each fermion $\alpha,\beta \in \{e,f\}$ is a hybrid of a trion and a hole.

The matrix structure of the problem for the free actions is
\begin{align}
  \label{eq:5}
  S_{b,0} &= \int_{-\infty}^{+\infty} \text
    dt \int_{\bs k} (\overline b^{\rm cl}_{\bs k}, \overline b^{\rm q}_{\bs k})
    \begin{pmatrix}
      0 & [G_{b,0}^{-1}]^{\rm A}\\
  [G_{b,0}^{-1}]^{\rm R}& [G_{b,0}^{-1}]^{\rm K}
    \end{pmatrix}
    \begin{pmatrix}
      b^{\rm cl}_{\bs k}\\
      b^{\rm q}_{\bs k}
    \end{pmatrix},\\
  S_{\alpha,0} &=  \int_{-\infty}^{+\infty} \text
    dt \int_{\bs p}  (\overline \alpha^{\rm r}_{\bs p}, \overline \alpha^{\rm a}_{\bs p})
    \begin{pmatrix}
    [G_{\alpha,0}^{-1}]^{\rm R}   & [G_{\alpha,0}^{-1}]^{\rm K}\\
  0 & [G_{\alpha,0}^{-1}]^{\rm A}
    \end{pmatrix}
    \begin{pmatrix}
      \alpha^{\rm r}_{\bs p}\\
      \alpha^{\rm a}_{\bs p}
    \end{pmatrix},
\end{align}
where in particular the Keldysh propagators
\begin{subequations}
    \begin{align}
        G_{b,0}^{\rm K}(\bs k,E) 
        &=  -i 2\pi F_b(E)\,\delta(E-\omega({\bs k})),\\
G_{\alpha,0}^{\rm K}(\bs p,E)  &= -i 2\pi F_\alpha(E)\, \delta(E-\lambda_{\alpha}(\bs p)) \qquad \alpha \in \{e,f\}
    \end{align}
\end{subequations}
 define the density functions $F_b,F_e,F_f$.

The interaction term Eq.~\eqref{eq:7} entails a nonzero self-energy which appears in Dyson's equation,
\begin{align}\label{eq:dyson}
\bs G_i(t,t')&= \bs G_{i,0}(t-t') 
 +\iint \text dt_1 \text dt_2 \bs G_{i,0}(t-t_1)\bs \Sigma_i(t_1,t_2) \bs G_{i}(t_2-t'),
\end{align}
for each particle species $i \in \{b,e,f\}$. In this expression, boldface signals the Keldysh matrix structure, and momentum indices are kept implicit. 
To the second order in interaction coefficients $g_{\alpha \beta}^{(\pm)}$,
we obtain the following self energies, defined from the same matrix structure as $\bs G_{i,0}^{-1}$ above. For fermions $\alpha \in \{e,f\}$,
\begin{align}
\Sigma^{\rm R/A}_\alpha(\bs X,t;\bs p, E) 
&=  \frac {1}{2} \int_{\bs k} |g^{(-)}_{\alpha\alpha}(\bs p-\bs k;\bs k)|^2 \bigg(\frac {  F_b(\omega_{\bs k}) +   F_\alpha(\lambda_{\alpha}(\bs p-\bs k))  }{  E - \omega_{\bs k} - \lambda_{\alpha}(\bs p-\bs k) \pm i 0 } \bigg )
+ |g^{(-)}_{\alpha\alpha}(\bs p;\bs k)|^2 \bigg (\frac {  F_b(\omega_{\bs k}) -    F_\alpha(\lambda_{\alpha}(\bs p+\bs k))  }{  E + \omega_{\bs k} - \lambda_{\alpha}(\bs p+\bs k) \pm i 0 }\bigg)\nonumber \\
&+  |g^{(-)}_{\alpha \ul \alpha}(\bs p-\bs k;\bs k)|^2 \bigg(\frac {  F_b(\omega_{\bs k}) +   F_{\ul \alpha}(\lambda_{\ul \alpha}(\bs p-\bs k))  }{  E - \omega_{\bs k} - \lambda_{\ul \alpha}(\bs p-\bs k) \pm i 0 }\bigg)
+|g^{(-)}_{\ul \alpha \alpha}(\bs p;\bs k)|^2 \bigg(\frac {  F_b(\omega_{\bs k}) -    F_{\ul \alpha}(\lambda_{\ul \alpha}(\bs p+\bs k))  }{  E + \omega_{\bs k} - \lambda_{\ul \alpha}(\bs p+\bs k)  \pm i 0 }\bigg) ,\\
\Sigma^{\rm K}_\alpha(\bs X,t;\bs p, E)  
&=  -i\pi  \int_{\bs k} |g^{(-)}_{\alpha\alpha}(\bs p-\bs k;\bs k)|^2 
\Big( - \big ( 1 + F_b(\omega_{\bs k})   F_\alpha(\lambda_{\alpha}(\bs p-\bs k)) \big ) \,\delta(E-\omega_{\bs k} -\lambda_{\alpha}(\bs p-\bs k) ) \bigg ) \nonumber \\ 
& + |g^{(-)}_{\alpha\alpha}(\bs p;\bs k)|^2 \bigg ( \big (1-F_b(\omega_{\bs k}) F_\alpha(\lambda_{\alpha}(\bs p + \bs k)) \big )\, \delta(E-\omega_{\bs k}+\lambda_{\alpha}(\bs p + \bs k)) \Big ) \nonumber \\
& +|g^{(-)}_{\alpha \ul \alpha}(\bs p-\bs k;\bs k)|^2
\Big(- \big ( 1+F_b(\omega_{\bs k}) 
F_{\ul \alpha}(\lambda_{\ul \alpha}(\bs p-\bs k))  \big )\, 
\delta(E-\omega_{\bs k}-\lambda_{\ul \alpha}(\bs p-\bs k))\Big)\nonumber\\
&+ |g^{(-)}_{\ul \alpha \alpha}(\bs p;\bs k)|^2 
\Big( \big (1-F_b(\omega_{\bs k})  F_{\ul \alpha}(\lambda_{\ul \alpha}(\bs p+\bs k)) \big )\,
\delta(E+\omega_{\bs k}-\lambda_{\ul \alpha}(\bs p+\bs k))\Big) ,
\end{align}
where $\alpha = e \Rightarrow {\ul \alpha} = f$ and conversely, and ``R'' corresponds to $+i0$ and ``A'' to $-i0$. 
For bosons, similarly,
  \begin{align}
  \label{eq:21}
\Sigma^{\rm R/A}_b(\bs X,t;\bs k, E)
&=  \frac {1}{2} \int_{\bs p} \bigg\{ \sum_{\alpha \in \{c,d\}} 
|g^{(-)}_{\alpha\alpha}(\bs p;\bs k)|^2 
\frac {  F_\alpha(\lambda_{\alpha}(\bs p+\bs k)) - F_\alpha(\lambda_{\alpha}(\bs p))  }{  E - \lambda_{\alpha}(\bs p+\bs k) + \lambda_{\alpha}(\bs p) \pm i 0 }
\nonumber \\& 
+  \sum_{\alpha \in \{c,d\}} |g_{\alpha \ul \alpha}^{(-)}(\bs p;\bs k)|^2 \frac {  F_\alpha(\lambda_{\alpha}(\bs p+\bs k)) -    F_{\ul \alpha}(\lambda_{\ul \alpha}(\bs p))  }{  E - \lambda_{\alpha}(\bs p+\bs k) + \lambda_{\ul \alpha}(\bs p) \pm i 0 }\bigg\},\\
\Sigma^{\rm K}_b(\bs X,t;\bs k, E) 
&= -i\pi\int_{\bs p}
 \bigg\{  \sum_{\alpha \in \{c,d\}} 
 |g_{\alpha\alpha}^{(-)}(\bs p;\bs k)|^2 \Big ( 1 -F_\alpha(\lambda_{\alpha}(\bs p+\bs k)) F_\alpha(\lambda_\alpha(\bs p))  \Big )
 \delta \big ( E  -\lambda_{\alpha}(\bs p+\bs k) +\lambda_{\alpha}(\bs p) \big ) \nonumber \\
 &+ \sum_{\alpha \in \{c,d\}} 
 |g_{\alpha \ul \alpha}^{(-)}(\bs p;\bs k)|^2 
 \Big ( 1 -F_\alpha(\lambda_\alpha(\bs p+\bs k)) F_{\ul \alpha}(\lambda_{\ul \alpha}(\bs p))  \Big )
 \delta \Big ( E  -\lambda_{\alpha}(\bs p+\bs k) +\lambda_{\ul \alpha}(\bs p) \Big )\nonumber \bigg\} .
\end{align}

In terms of these self-energies, the collision integral for particle species $\alpha\in \{e,f\},b$ reads
\begin{subequations}
\begin{align}
\label{eq:IfromSigma}
I_\alpha(\bs p) &= i \Sigma^{\rm K}_\alpha(\bs p, \lambda_\alpha(\bs p)) + 2 {\rm Im}\Sigma^{\rm R}_\alpha(\bs p, \lambda_\alpha(\bs p)) F_\alpha(\bs p,\lambda_\alpha(\bs p)),\\
I_b(\bs k) &= i \Sigma^{\rm K}_b(\bs k, \omega_{\bs k}) + 2 {\rm Im}\Sigma^{\rm R}_b(\bs k, \omega_{\bs k}) F_b(\bs k,\omega_{\bs k}).
\end{align}
\end{subequations}

Explicitly, 
\begin{align}
I_f  (\bs p) &= \pi \int_{\bs k} \delta(\lambda_{f}(\bs p)-\omega_{\bs k}-\lambda_{f}(\bs p-\bs k)) \, |g_{ff}^{(-)}(\bs p-\bs k;\bs k)|^2 \, [1+ F_b (\omega_{\bs k}) F_f (\lambda_f (\bs p-\bs k )) -( F_b (\omega_{\bs k})+ F_f (\lambda_f (\bs p -\bs k))) F_f (\lambda_f (\bs p ))] \nonumber\\
&+\delta(\lambda_{f}(\bs p)+\omega_{\bs k}-\lambda_{f}(\bs p+\bs k))\, |g_{ff}^{(-)}(\bs p;\bs k)|^2 \, [1- F_b (\omega_{\bs k}) F_f (\lambda_f (\bs p+\bs k )) +( F_b (\omega_{\bs k })- F_f (\lambda_f (\bs p+\bs k ))) F_f (\lambda_f (\bs p ))]) \nonumber\\
&+ \delta(\lambda_{f}(\bs p)-\omega_{\bs k}-\lambda_{e}(\bs p-\bs k))\, |g_{fe}^{(-)}(\bs p-\bs k;\bs k)|^2\,
[1+ F_b (\omega_{\bs k}) F_e (\lambda_e (\bs p-\bs k )) -( F_b (\omega_{\bs k})+ F_e (\lambda_e (\bs p -\bs k))) F_f (\lambda_f (\bs p ))]) \nonumber\\
&+\delta(\lambda_{f}(\bs p)+\omega_{\bs k}-\lambda_{e}(\bs p+\bs k)) \,|g_{ef}^{(-)}(\bs p;\bs k)|^2\,[1- F_b (\omega_{\bs k }) F_e (\lambda_e (\bs p + \bs k)) +( F_b (\omega_{\bs k })- F_e (\lambda_e (\bs p +\bs k))) F_f (\lambda_f (\bs p ))])
\end{align}
and $I_e(\bs p)$ is obtained similarly by just exchanging $f \leftrightarrow e$, and
\begin{align}
I_b (\bs k) &= \pi \int_{\bs p}    \delta ( \omega_{\bs k}  -\lambda_{f}(\bs p+\bs k) +\lambda_{f}(\bs p) ) \,|g_{ff}^{(-)}(\bs p ; \bs k)|^2 \,
[   ( 1 -F_f(\lambda_f (\bs k+\bs p )) F_f(\lambda_f (\bs p ))    ) +   (  F_f(\lambda_f (\bs p )) -  F_f(\lambda_f (\bs p+\bs k)   ) )   F_b(\omega_{\bs k})    ] \nonumber\\
 &+ \delta ( \omega_{\bs k}  -\lambda_{e}(\bs p+\bs k) +\lambda_{e}(\bs p) ) \, |g_{ee}^{(-)}(\bs p ; \bs k)|^2 \,
  [   ( 1 -F_e(\lambda_e (\bs k+\bs p )) F_e(\lambda_e (\bs p ))    ) +   (  F_e(\lambda_e (\bs p )) -  F_e(\lambda_e (\bs p+\bs k)   ) ) F_b(\omega_{\bs k})    ]\nonumber \\
  &+ \delta ( \omega_{\bs k}  -\lambda_{f}(\bs p+\bs k) +\lambda_{e}(\bs p) ) \,|g_{fe}^{(-)}(\bs p ; \bs k)|^2 \,
    [   ( 1 -F_f(\lambda_f (\bs k+\bs p )) F_e(\lambda_e (\bs p ))    ) +   (  F_e(\lambda_e (\bs p )) -  F_f(\lambda_f (\bs p+\bs k)   ) ) F_b(\omega_{\bs k})    ] \nonumber\\
&  + \delta ( \omega_{\bs k}  -\lambda_{e}(\bs p+\bs k) +\lambda_{f}(\bs p) )\,  |g_{ef}^{(-)}(\bs p ; \bs k)|^2 \,
     [   ( 1 -F_e(\lambda_e (\bs k+\bs p )) F_f(\lambda_f (\bs p ))    ) +   (  F_f(\lambda_f (\bs p )) -  F_e(\lambda_e (\bs p+\bs k)   ) ) F_b(\omega_{\bs k})    ] .
\end{align}
These are the quantities appearing on the right-hand side of Eqs.~\eqref{eqs:boltzmann}.

\section{Solution of the coupled Boltzmann's equations in a magnetic field}

At this stage, one has obtained a system of three coupled Boltzmann equations, for the three species of particles, with unknowns ${\sf g}_\alpha(\bs p),{\sf g}_b(\bs k)$. The system of equations, even when keeping only the expansion of collision integrals to linear order in out-of-equilibrium  distributions ${\sf g}_i, i\in \{b,e,f\}$, is not diagonal in momentum space. The equations for the fermion species may then be written in compact matrix notation, in particular
\begin{align}
\label{eq:25}
   - (1/2) \, I_{\alpha}(\mb p ) = M^\alpha_{\mb p, \mb q} \,{\sf g}_\alpha ({\mb q}) + N^{\alpha}_{\mb p, \mb q}\, {\sf g}_{\ul \alpha}({ \mb q})+ C^{\alpha}({\mb q}),
\end{align} 
with implicit summation (i.e. integration) over repeated momentum indices. 
Here $C^{\alpha}({\mb p}) = O({\sf g}_b)$. Because bosons are put out of equilibrium by their coupling to the fermions, ${\sf g}_b=O(|g|^2)$ whence $C^{\alpha}({\mb p}) = O(|g|^4)$: this term is higher-order in perturbation than the other terms appearing in Eq.~\eqref{eq:25}, and we neglect it in the following.

Explicitly,
\begin{align}
    M^{f}_{\mb p, \mb q}
    &= -2\pi \delta_{\mb p, \mb q} \int_{\mb k} \bigg\{ |g^{(-)}_{ff}(\mb p-\mb k;\mb k)|^2 \, \delta\big(\omega_{\mb k}+\lambda_{f}( \mb p-\mb k )- \lambda_{f}( \mb p )\big)
 \big(n_{\rm B} (\omega_{\mb k}) - n_{\rm F} (\lambda_{f}(\mb p - \mb k)) + 1\big)  \nonumber\\ 
   &\qquad  \qquad\qquad + |g^{(-)}_{ff}(\mb p;\mb k)|^2 \, \delta\big(-\omega_{\mb k}+\lambda_{f}( \mb p+\mb k )- \lambda_{f}( \mb p )\big) \big( n_{\rm B} (\omega_{\mb k}) + n_{\rm F} (\lambda_{f}(\mb p + \mb k)) \big) \nonumber \\
   &\qquad \qquad \qquad +
  |g^{(-)}_{fe}(\mb p-\mb k;\mb k)|^2\delta\big(\omega_{\mb k}+\lambda_{e}(\mb p - \mb k )- \lambda_{f}( \mb p )\big)
  \big(n_{\rm B} (\omega_{\mb k}) - n_{\rm F} (\lambda_{e}(\mb p -\mb k)) + 1\big)  \nonumber\\
  & \qquad \qquad\qquad +|g^{(-)}_{ef}(\mb p,\mb k)|^2 \delta\big(-\omega_{\mb k}+\lambda_{e}(\mb p - \mb k )- \lambda_{f}(\mb p ) \big) 
    \big( n_{\rm B} (\omega_{\mb k}) + n_{\rm F} (\lambda_{e}( \mb p - \mb k)) \big) \bigg\} \nonumber \\
 &+  2\pi \bigg\{ |g^{(-)}_{ff}(\mb q;\mb q-\mb p)|^2  \delta\big(\omega_{\mb q-\mb p}+\lambda_{f}( \mb q )- \lambda_{f}( \mb p )\big)
 \big( n_{\rm B} (\omega_{\mb q - \mb p}) + n_{\rm F} (\lambda_{f}( \mb p)) \big) \nonumber \\
 &\qquad +|g^{(-)}_{ff}(\mb p;\mb q-\mb p)|^2
 \delta\big(-\omega_{\mb q- \mb p}+\lambda_{f}(\mb q) - \lambda_{f}( \mb p )\big)
\big(n_{\rm B} (\omega_{\mb q-\mb p}) - n_{\rm F} (\lambda_{f}(\mb p)) + 1\big) \bigg \},\\
  N^{f}_{\mb p , \mb q} &= 2\pi  \Big \{
    |g^{(-)}_{fe}(\mb q; \mb p -\mb q)|^2  \delta\big(\omega_{\mb p- \mb q}+\lambda_{e, \mb q }- \lambda_{f, \mb p }\big)
     \big( n_{\rm B} (\omega_{\mb p-\mb q}) + n_{\rm F} (\lambda_{f, \mb p}) \big)   \nonumber\\
 &\qquad + |g^{(-)}_{ef}(\mb p;\mb q - \mb p)|^2 \, \delta\big(-\omega_{\mb q- \mb p}+\lambda_{e, \mb q }- \lambda_{f, \mb p }\big)
    \big(n_{\rm B} (\omega_{\mb q-\mb p}) - n_{\rm F} (\lambda_{f,\mb p}) + 1\big)  \Big \} ,
\end{align}
and similarly $M^e_{\mb p,\mb q}, N^e_{\mb p,\mb q}$ are obtained by exchanging $f \leftrightarrow e$.
The set of two coupled Boltzmann equations for the two species $(e,f)$ of fermions then reads (cf Eq.~\eqref{eq:lhs-in-field})
\begin{align}
\label{eq:30}
\partial_t {\sf g} + \xi {\sf e}(\bs E \cdot \bs v) n'_{\rm F}(\lambda) 
- {\sf e}(\bs B \times \bs v)\cdot \partial_{\bs p} {\sf g} 
= - {\sf g}/{\tau} +\begin{bmatrix}
        M^e &N^e \\N^f & M^f 
    \end{bmatrix} {\sf g}  ,
\end{align}
where all objects ${\sf g},\xi, \bs v,\lambda$ should be understood as infinite vectors indexed by momentum $\bs p$ and two-flavor index $(e,f)$.
Notice that we have added an extra relaxation term $-{\sf g}/\tau$, in a relaxation time approximation fashion, accounting for extrinsic effects due for instance to disorder. In the following we will take $\partial_t {\sf g} \rightarrow 0$ for clarity, but note that time dependence can be restored at any stage by replacing $1/\tau \rightarrow 1/\tau + \partial_t$.

To solve Eq.~\eqref{eq:30} we introduce the Ansatz ${\sf g}_\alpha(\bs p)=- \bs v_\alpha(\bs p) \cdot \bs K_{\alpha}(\lambda_\alpha(\bs p))$ for $\alpha \in \{e,f\}$.  Using the fact that the two species' velocities are collinear to each other, $\bs v_\alpha \equiv \eta_{\ul\alpha\alpha}\bs v_{\ul \alpha}$ with $\eta_{ef}\eta_{fe}=1$, and that the $M^\alpha_{\mb{pq}},N^\alpha_{\mb{pq}}$ matrix elements are sizeable for momenta $\mb p \approx \mb q$ only, we obtain the equation
\begin{equation}
\label{eq:s31}
    {\sf e}{\sf E}_\alpha - ({\sf e}/m_\alpha){\sf B}_\alpha \times \bs K_\alpha - {\Gamma}_\alpha \bs K_\alpha = 0,
\end{equation}
having defined the auxiliary quantities ${\sf B}_\alpha = \bs B$ 
and $\Gamma_\alpha = 1/\tau - M^\alpha$ and ${\sf E}_\alpha =  n'_{\rm F}(\lambda_\alpha)\xi_\alpha \bs E + N^\alpha \eta_{\alpha{\ul \alpha}} \bs K_{\ul \alpha} $. 

This has a solution
\begin{equation}
\label{eq:s32}
  \bs K_{\alpha}
    = {\sf e} \Big [ \Gamma_\alpha^2 + ({\sf e}/m_\alpha)^2 {\sf B}_\alpha^2 \Big ]^{-1}\Big ( \Gamma_\alpha  + ({\sf e}/m_\alpha){\sf B}_\alpha \!\times \Big ){\sf E}_\alpha ,
\end{equation}
where we recall that ${\sf E}_\alpha$ depends on $\bs K_{\ul \alpha}$. 
Plugging Eq.~\eqref{eq:s32} ($\ul \alpha$) back into Eq.~\eqref{eq:s31} ($\alpha$), one obtains the equation
\begin{equation}
\label{eq:s33}
     {\sf e}\tilde {\sf E}_\alpha - ({\sf e}/m_\alpha)\tilde {\sf B}_\alpha \times \bs K_\alpha - \tilde {\Gamma}_\alpha \bs K_\alpha = 0,
\end{equation}
having defined the new auxiliary quantities
\begin{subequations}
\begin{align}
 \tilde \Gamma_\alpha 
 &= \Gamma_\alpha - N_\alpha \Big [ \Gamma_{\ul \alpha}^2 + ({\sf e}/m_{\ul \alpha})^2 {\sf B}_{\ul \alpha}^2 \Big ]^{-1}\ \Gamma_{\ul \alpha} N_{\ul \alpha} ,\\
\tilde {\sf B}_\alpha 
&= \bs B \Big ( 1 + N_\alpha \Big [ \Gamma_{\ul \alpha}^2 + ({\sf e}/m_{\ul \alpha})^2 {\sf B}_{\ul \alpha}^2 \Big ]^{-1} N_{\ul \alpha} (m_\alpha/m_{\ul \alpha}) \Big ),\\
\tilde {\sf E}_\alpha 
&= n'_{\rm F}(\lambda_\alpha)\xi_\alpha \bs E
+ N_\alpha \eta_{\alpha{\ul \alpha}} \Big [ \Gamma_{\ul \alpha}^2 + ({\sf e}/m_{\ul \alpha})^2 {\sf B}_{\ul \alpha}^2 \Big ]^{-1} \Big ( \Gamma_{\ul \alpha}  + ({\sf e}/m_{\ul \alpha}){\sf B}_{\ul \alpha} \!\times \Big )\bs E\,  n'_{\rm F}(\lambda_{\ul \alpha})\xi_{\ul \alpha} ,
    \end{align}
\end{subequations}
to be understood as infinite matrices ($\tilde \Gamma,\tilde {\sf B}$) and vectors ($\tilde {\sf E}$) indexed by momenta. This Eq.~\eqref{eq:s33} can be solved similar to the above, whence the final result
\begin{equation}
  {\sf g}_\alpha(\bs p)
  = - {\sf e}\bs v_\alpha(\bs p) \cdot  \Big [ \tilde \Gamma_\alpha^2 + ({\sf e}/m_\alpha)^2 \tilde {\sf B}_\alpha^2 \Big ]^{-1}_{\bs p,\bs l}\Big ( \tilde \Gamma_\alpha  + ({\sf e}/m_\alpha)\tilde {\sf B}_\alpha \!\times \Big )_{\bs l,\bs q}\tilde {\sf E}_\alpha(\bs q) .
\end{equation}
This is the off-equilibrium distribution involved in the current, Eq.~\eqref{eq:Jh} that we rewrite here as
\begin{equation}
   \bs J_h = {\sf e} \sum_{\alpha \in \{e,f\}} \int_{\bs p} {\sf g}_\alpha(\bs p)\, \xi_\alpha
    \frac{\partial \epsilon_h}{\partial \bs p} .
   \end{equation}

\bibliography{library}